\documentclass[conference,a4paper]{APSIPA2026}

\usepackage{amsmath}
\usepackage{amsfonts}
\usepackage{amssymb}

\usepackage{graphicx}
\usepackage{multirow}
\usepackage{threeparttable}
\usepackage{tabularx}
\usepackage{booktabs}

\usepackage{comment}
\usepackage{xurl}

\usepackage[
    backend=biber,
    style=ieee
]{biblatex}
\usepackage{geometry}
\usepackage{fancyhdr}

\fancypagestyle{firststyle}{
    \fancyhf{}
    \fancyhead[C]{%
        2026 Asia Pacific Signal and Information Processing Association
        Annual Summit and Conference (APSIPA ASC)%
    }
}

\usepackage{hyperref}

\begin{document}

\title{Exploring Efficient Waveform Diffusion Models \\ for Foley Sound Generation}

\author{
\authorblockN{
Runwu Shi\authorrefmark{1},
Chang Li\authorrefmark{2},
Jiahui Li\authorrefmark{1},
Jiang Wang\authorrefmark{1},
Yaozhong Kang\authorrefmark{1},
Nabeela Khan\authorrefmark{1},
Linghan Fang\authorrefmark{3},\\
Benjamin Yen\authorrefmark{1},
Takeshi Ashizawa\authorrefmark{1},
Kazuhiro Nakadai\authorrefmark{1}
}

\authorblockA{
\authorrefmark{1}
Institute of Science Tokyo, Tokyo, Japan}

\authorblockA{
\authorrefmark{2}
University of Science and Technology of China,
\authorrefmark{3}
Technical University of Munich \\
E-mail: shirunwu@ra.sc.e.titech.ac.jp}
}

\maketitle
\thispagestyle{firststyle}
\pagestyle{empty}

\begin{abstract}
Recent advances in diffusion models have enabled high-fidelity Foley sound generation directly in the waveform space. Existing waveform diffusion models primarily rely on time-domain architectures, such as CNN-based U-Nets and DiffWave-style models, or frequency-domain Transformers modeling temporal dependencies. However, these systems are typically built with large model capacities and substantial computational costs, leaving compact and efficient waveform diffusion architectures largely underexplored. In this work, we introduce a Dual-Path (DP) architecture for waveform diffusion that performs dimension-wise self-attention along both subband and frame axes in the time–frequency domain. This DP design enables fine-grained temporal–spectral modeling while maintaining high efficiency. Based on the proposed DP backbone, we develop two variants: DP-DiT and DP-U-Net. Experiments on the DCASE and FSD-Kaggle2018 datasets demonstrate their superior performance. Notably, the 3M parameter variant achieves performance comparable to models with more than 50M parameters. Audio samples are available at \href{https://samplesdemo.github.io/DP-Foley/}{https://samplesdemo.github.io/DP-Foley/}.
\end{abstract}

\section{Introduction}
Foley sound generation aims to synthesize sound effects that are tightly aligned with events and timing, requiring both high perceptual fidelity and accurate fine-grained temporal structure. A variety of generative paradigms have been explored for this task, including Generative Adversarial Networks (GANs) \cite{ghose2022foleygan}, autoencoders \cite{xie2023x}, and diffusion models \cite{qi2024mtdiffusion, huang2025rhythmic}. In this work, we focus on raw audio generation and adopt diffusion models to generate high-quality Foley sound.

Many diffusion-based methods for audio generation have been proposed, targeting diverse output spaces. These approaches can be broadly categorized into three classes: (1) waveform-level generation, which directly synthesizes raw audio signals \cite{chung2024t}, (2) latent-space generation, which operates in a compressed representation space commonly learned by an autoencoder \cite{yuan2023text}, and (3) acoustic-feature generation, which produces intermediate representations such as mel-spectrograms that are later decoded into waveforms through vocoders \cite{liu2023audioldm}. In this paper, we adopt waveform-level generation, motivated by both fidelity and efficiency considerations. Compared to latent-space or acoustic-feature approaches, waveform-level generation enables end-to-end modeling without relying on intermediate representations and separate decoders, thereby offering a more compact formulation. Moreover, with appropriate architectural design, waveform models can directly capture transient-rich and fine-grained temporal--spectral structures that are crucial for producing convincing Foley sounds.

At the same time, it is worth noting that recent large-scale audio and music generation systems often operate in compressed domains (e.g., neural audio codec tokens or learned latents) \cite{liu2024audioldm}, largely because raw waveforms are extremely high-dimensional and sequences at audio sampling rates are prohibitively long to model at scale. Rather than addressing the complexity through model scaling or domain compression, in this work, we investigate the limits of direct waveform diffusion under strict parameters and compute budgets. In this context, we study lightweight architecture paradigms that enable raw waveform synthesis with only a few million parameters while maintaining competitive perceptual quality, prioritizing efficiency, deployability, and end-to-end fidelity. Nevertheless, existing waveform-based diffusion backbones (e.g., CNN U-Nets and DiffWave-style architectures) often rely on very deep networks and relatively large parameter counts to model high-dimensional raw signals. Meanwhile, fully self-attention backbones working in the time domain remain costly, and recent alternatives that apply temporal attention in the time--frequency (TF) domain to predict complex spectrograms to alleviate the long sequence burden~\cite{ku2025generative}, yet they typically emphasize {global temporal} modeling and may under-exploit fine-grained Intra-frame and Intra-frequency structures that are crucial for expressive Foley synthesis.

\begin{figure}[!tb]
  \centering
  \centerline{\includegraphics[width=8.5cm]{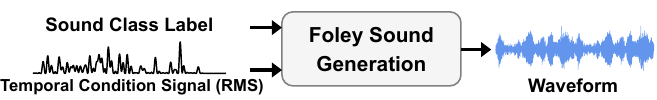}}
%  \vspace{2.0cm}
  % \centerline{(a) Result 1}\medskip
\caption{Class and temporal signal guided waveform generation.}
\label{fig:one}
\vspace{-1em} 
\end{figure}
% \section{Related Works}
\begin{figure*}[!t]
  \centering
  \includegraphics[width=\textwidth]{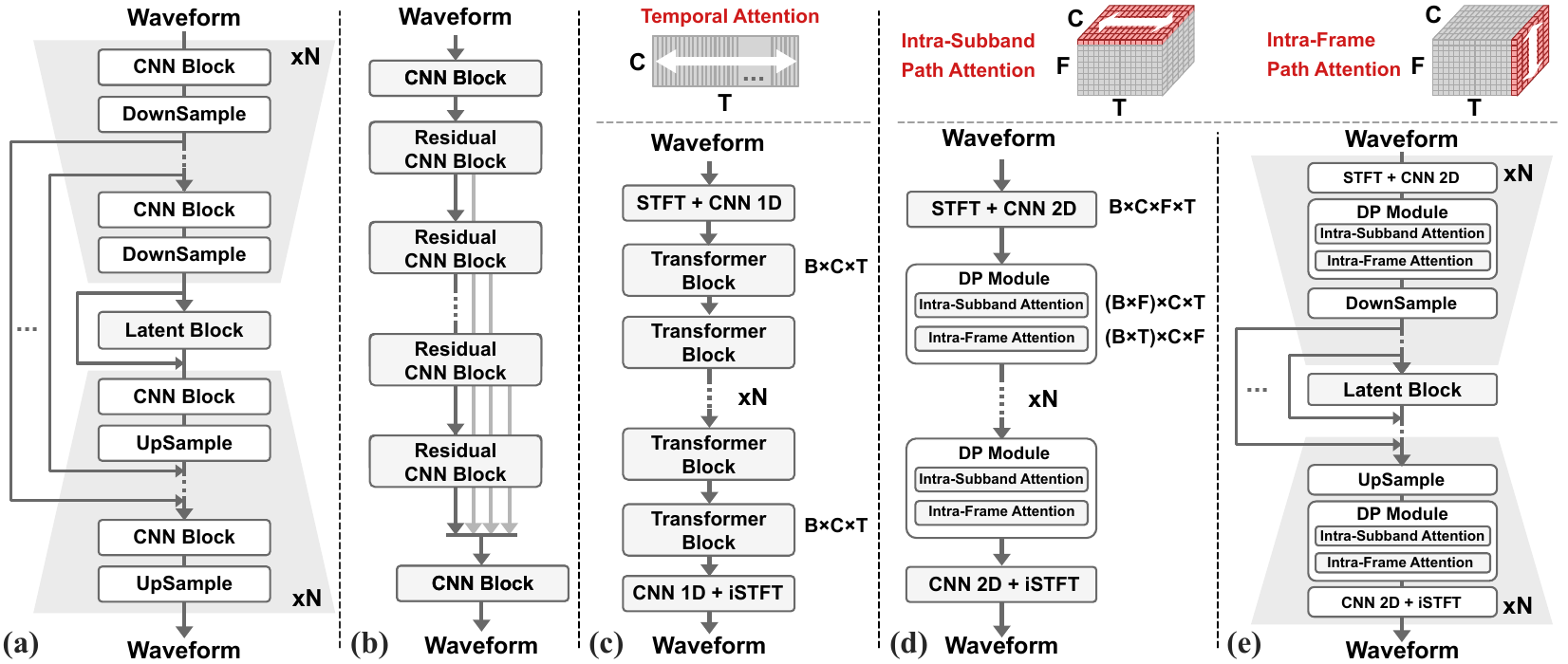}
\caption{Five model variants for raw waveform generation: (a) CNN U-Net, (b) DiffWave, (c) TF temporal-attention DiT (TF-DiT), (d) TF Dual Path-attention DiT, (e) TF Dual Path-attention U-Net. Conditioning signals (class embedding, diffusion time step embedding, and RMS energy signal) are omitted for clarity.}
  \label{fig:two}
  \vspace{-1em}
\end{figure*}

Motivated by the fine-grained frame-path and subband-path modeling paradigm that has proven effective in speech separation and enhancement literature~\cite{luo2020dual}, we propose a \textbf{Dual-Path (DP)} backbone that performs attention-based modeling along both the Intra-frame and Intra-subband axes in the TF domain. Building upon this DP module, we instantiate two lightweight diffusion backbones: \textbf{DP-DiT} (DP Diffusion Transformer) and \textbf{DP-U-Net}, both of which support class and temporal energy signal conditioning, as shown in Figure~\ref{fig:one}. 

We conduct experiments on DCASE and FSD-Kaggle2018 datasets, and the results demonstrate that our DP-based models consistently outperform state-of-the-art baselines across all objective and subjective metrics. Notably, our architectures achieve superior performance with significantly fewer parameters. This demonstrates that high-quality waveform diffusion can be realized at a remarkably small scale through optimized design, providing a resource-efficient yet high-fidelity solution for deployable Foley generation.

\section{Related Works}
Diffusion models that directly operate on raw waveforms have been widely explored for various audio tasks, e.g., audio generation, speech enhancement, and source separation~\cite{lu2021study, shi2026unsupervised, colombo2025mambafoley}. To support direct waveform synthesis, several diffusion-based backbones have been developed in either the time or frequency domain, as summarized in Figure~\ref{fig:two}. We broadly group them into five categories: CNN U-Net, DiffWave-style, TF temporal-attention DiT, TF DP-attention DiT, and TF DP-attention U-Net. CNN U-Nets cover most Foley waveform generators: T-Foley~\cite{chung2024t} adopts a CNN U-Net and introduces Block-FiLM for temporal control using the RMS energy, while Mamba-Foley~\cite{colombo2025mambafoley} augments a similar backbone with a Mamba unit at the bottleneck; both build upon the U-Net design popularized by DAG~\cite{pascual2023full}. DiffWave~\cite{kongdiffwave} follows a WaveNet-inspired time-domain architecture~\cite{van2016wavenet}, which has been widely applied for vocoder tasks \cite{nguyen2024fregrad}. In the frequency domain,  Undiff \cite{iashchenko2023undiff} proposes FFC-AE, a convolutional network operating on complex-valued Short-time Fourier transform (STFT) spectrograms, and CQT-Diff, a CNN U-Net that operates in the CQT domain~\cite{moliner2023solving}. There are also transformer-based methods, some of which add the self-attention mechanism into CNN U-Net backbones, and \cite{ku2025generative} uses a transformer to predict complex STFT coefficients as temporal sequences.

\section{Methodology}
\subsection{Overall Model Architecture}
We design two backbone paradigms, a DiT-style model and a U-Net-style model, as illustrated in Figure~2(d,e). 
Both architectures share the same core module: DP Module equipped with dimension-wise self-attention, including Intra-subband and Intra-frame attention, as detailed in the next subsection. 
Accordingly, we name our models DP-DiT and DP-U-Net. DP-DiT processes the feature map at the original TF resolution without any downsampling. In contrast, DP-U-Net follows a five-stage encoder-decoder structure comprising two downsampling stages, a latent bottleneck stage, and two symmetric upsampling stages. Both backbones share the same TF front-end and waveform generation head. The input waveform $x$ is transformed into a complex STFT spectrogram with size $2\times F\times T$, where $F$ and $T$ denote the number of frequency bins and time frames, respectively. A 2D convolutional projection then maps this spectrogram to a $C$-channel feature map of size $C\times F\times T$ for subsequent processing. After being processed by the DP-DiT or DP-U-Net, the features are projected back and converted to the waveform domain via a transposed convolutional layer and inverse STFT, yielding the predicted noise $\boldsymbol{\epsilon}$. Specifically, DP-U-Net adopts 2D shuffle-based upsampling and downsampling mechanisms. Utilizing PixelShuffle and PixelUnshuffle layers combined with convolutions, DP-U-Net scales the spatial resolution by a factor of 2 at each stage to capture multi-scale features.

\begin{figure}[!tb]
  \centering
  \centerline{\includegraphics[width=8cm]{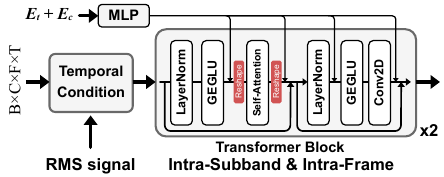}}
 % \vspace{2.0cm}
  % \centerline{(a) Result 1}\medskip
\caption{Diagram of the DP module. The transformer block is applied twice (Intra-subband and Intra-frame), shown once for clarity. $E_t$ and $E_c$ mean diffusion step and class embedding.}
\label{fig:3}
\vspace{-2em} 
\end{figure}

\subsection{DP block with dimension-wise self-attention}
Figure 3 illustrates the architecture of the DP Module. This module begins with an optimized Block-FiLM mechanism for temporal conditioning \cite{choi2023foley}. Departing from the original cascaded two-layer design, which employs two consecutive FiLM-and-convolution blocks. We modify the structure into a single-layer configuration per module. Despite the reduction in depth, we align the operational resolution of Block-FiLM with the intermediate TF representation. This enables fine-grained, TF-bin-level control that provides superior temporal control accuracy. For class and diffusion step conditioning, each module incorporates a diffusion timestep embedding $E_t$ and a learnable sound class embedding $E_c$ via the AdaLN-Zero mechanism \cite{peebles2023scalable}. This mechanism predicts scaling and shifting parameters from the combined embedding $E_t + E_c$ for modulation. After this, given the feature map in $\mathbb{R}^{B\times C\times F\times T}$, where $B$ is the batch size, the module captures joint dependencies by alternating between Intra-subband and Intra-frame attention. Specifically, the Intra-subband path reshapes the features into $\mathbb{R}^{(B\times F)\times C\times T}$ to model temporal dynamics, while the Intra-frame path reshapes them into $\mathbb{R}^{(B \times T)\times C\times F}$ to capture instantaneous spectral relationships. Each stage follows a Pre-Norm architecture, integrating multi-head self-attention and a feed-forward network containing GEGLU and Conv2D within a residual structure.

\begin{table*}[!t]
  \caption{Objective and subjective evaluation results. The best results are shown in bold. During testing, the RMS signal is extracted from the evaluation dataset. Real data's FAD is calculated between the training and evaluation datasets.}
  \label{tab:1}
  \centering
  \renewcommand{\arraystretch}{1.1}
  \small
  \begin{tabular}{l c | c c c | c c c | c c}
  % \toprule
    % & \multicolumn{2}{c|}{\textbf{Model Info}}
    & \multicolumn{1}{c}{}
    & \multicolumn{3}{c}{\textbf{DCASE (7 classes)}}
    & \multicolumn{3}{c}{\textbf{FSD-Kaggle2018 (41 classes)}}
    & \multicolumn{2}{c}{\textbf{Subjective evaluation}} 
    \\
    % \cmidrule(lr){1-3}
    % \cmidrule(lr){4-6}
    % \cmidrule(lr){7-9}
  \toprule
    \textbf{Method} 
    & \textbf{Params}
    & \textbf{E-L1} $\downarrow$ & \textbf{FAD-P} $\downarrow$ & \textbf{FAD-V} $\downarrow$
    & \textbf{E-L1} $\downarrow$ & \textbf{FAD-P} $\downarrow$ & \textbf{FAD-V} $\downarrow$ 
    & \textbf{Quality} $\uparrow$ & \textbf{Time Align} $\uparrow$
    \\
  \midrule
    Real Data 
    & -- 
    & -- & 18.82 & 3.95
    & -- & 69.71 & 8.41 
    & -- & --
    \\
  \midrule
    T-Foley \cite{chung2024t}
    &  74.09M
    & 0.035 & 34.97 & 10.11
    & 0.058 & 87.32 & 17.14 
    & 2.53 & 3.04 \\
    Mamba-Foley \cite{colombo2025mambafoley}
    & 58.81M
    & 0.021 & 29.65 & 6.03
    & 0.028 & 72.99 & 12.22 
    & 3.25 & 3.66 \\
    Diffwave \cite{kongdiffwave}
    & 12.58M
    & 0.019 & 52.00 & 11.06
    & 0.019 & 100.26 & 15.83 
    & 2.45 & 2.84 \\
    TF-DiT
    & 21.39M
    & 0.170 & 116.76 & 23.87
    & 0.110 & 130.66 & 32.83 
    & 0.79  & 0.82 \\
  \midrule
    \textbf{DP-DiT}
    & 3.27M
    & \textbf{0.009} & 30.07 & 7.02
    & 0.007 & 67.09 & 13.53 
    & 3.97 & 4.46 \\
    \textbf{DP-U-Net}
    & 8.57M
    & \textbf{0.009} & \textbf{27.48} & \textbf{5.34}
    & \textbf{0.006} & \textbf{62.77} & \textbf{11.55} 
    & \textbf{4.07} & \textbf{4.47} \\
    \textbf{DP-U-Net Small}
    & 3.26M
    & \textbf{0.009} & 27.98 & 6.34
    & 0.007 & 75.13 & 11.98 
    & 3.79 & 4.30 \\
    % DP-U-Net Large
    % & TF & 12.10M
    % & -- & -- & --
    % & 0.007 & 62.86 & 11.67 \\
  \bottomrule
  \end{tabular}
  \vspace{-1em} 
\end{table*}

\section{Experiments}
\subsection{Dataset}
We utilized two datasets to evaluate the proposed method. First is the Detection and Classification of Acoustic Scenes and Events (DCASE) task 7 dataset challenge \cite{choi2023foley}. This dataset contains 7 classes of sounds, including dog bark, footstep, gunshot, keyboard, vehicle, rainfall, and coughing. The developing dataset contains 4,850 audio samples, and the evaluation dataset contains 700 audio samples. We use the official development and evaluation datasets for training and evaluation. In addition, we utilize the FSD-Kaggle2018 dataset~\cite{fonseca2018general}, which comprises around 11,000 clips across 41 sound event categories. We use the provided training set with 9,470 clips for training and randomly choose 10 samples per class for testing. All audio samples are sampled at 22,050 Hz with a duration of 4 seconds. 

\subsection{Implementation Details}
For all DP-DiT and DP-U-Net variants, we compute the STFT using a window size of 510 and a hop length of 255. This yields a complex-valued TF representation of shape 352 $\times$ 256 (time $\times$ frequency bins) for each 4-second audio clip.
The temporal conditional RMS signal used for temporal control is extracted from the input waveform with a frame length of 512 and a hop length of 128. For DP-DiT, the channel dimension is set to $C=64$. The DP module is repeated 32 times, resulting in a total of 3.27M parameters. For DP-U-Net, we consider two model variants that share the same channel dimension $C=64$ but differ in the number of block repetitions. 
The standard DP-U-Net adopts a five-stage encoder--decoder layout with DP module repetitions of $\{1, 2, 2, 2, 1\}$, yielding 8.57M parameters. 
We further introduce a lightweight variant, {DP-U-Net Small}, which uses the same five-stage layout with a single DP module at each stage, resulting in only 3.26M parameters. We train all of the models for 500 epochs using the AdamW optimizer~\cite{loshchilov2017decoupled} with a learning rate of $1\times10^{-4}$. Denoising training is conducted for 200 steps with a linear noise schedule ranging from $10^{-4}$ to $2\times10^{-2}$. For sampling, we adopt the DDPM scheduler with classifier-free guidance (scale = 1.2), where 10\% of the conditions are randomly dropped during training for unconditional generation.

% For experiments conducted solely on the FSD-Kaggle2018 dataset, which contains a larger variety of sound types, we employ a higher-capacity model, \textit{DP-U-Net Large}, with block repetitions of $\{2, 3, 3, 3, 2\}$, resulting in 12.12M parameters.

\subsection{Compared Baselines}
We compare the proposed DP-DiT and DP-U-Net with several representative baselines, including T-Foley \cite{chung2024t}, Mamba-Foley \cite{colombo2025mambafoley}, DiffWave \cite{kongdiffwave}, and a TF-domain DiT model. T-Foley is a CNN-based U-Net with a recurrent unit in the latent stage for temporal modeling, for which we use the official pretrained checkpoint and original sampling settings\footnote{https://github.com/YoonjinXD/T-foley}. 
For Mamba-Foley, we implement it using the official codebase and train it from scratch under the same training settings as our methods\footnote{https://github.com/FurioColombo/mamba-foley}. 
Both models are substantially larger than ours, with 74.09M and 58.81M parameters, respectively. For the sampling setting, we use their original guidance scale. We also include DiffWave~\cite{kongdiffwave}, augmented with Block-FiLM conditioning for temporal control, configured with 24 residual blocks and a channel dimension of 98, resulting in 12.58M parameters. 
In addition, we compare with a TF-domain DiT baseline that adopts the same Transformer block as the proposed method, and predicts concatenated real and imaginary spectrogram components along the temporal dimension, as shown in Figure~\ref{fig:two} (c). The depth is set to 16, resulting in 21.39M parameters. 
Directly applying DiT in the time domain would be computationally expensive due to the high temporal resolution of feature maps produced by the 1D convolutional head,  which requires specifically designed downsampling mechanisms that are beyond the scope of this work, and thus a time-domain DiT baseline is not considered. For all methods, we use 200 sampling steps and adopt the same DDPM sampling strategy.

\subsection{Evaluation Metrics}
The evaluation includes both objective and subjective measures. For the objective evaluation, we use the Fréchet Audio Distance (FAD)~\cite{kilgour2018fr} to assess perceptual quality and the Event-L1 Distance (E-L1) to evaluate the temporal accuracy of generated sound, following previous works~\cite{chung2024t, colombo2025mambafoley}. The FAD is computed using VGGish (FAD-V) and PANNs-CNN14-32k (FAD-P) embeddings implemented in the \texttt{fadtk} toolkit.
% \footnote{https://github.com/DCASE2024-Task7-Sound-Scene-Synthesis/fadtk}. 
The E-L1 distance is computed on RMS energy feature of audio waveform, and is defined as $\text{E-L1}(E, \hat{E}) = \tfrac{1}{N}\sum_{i=1}^{N}\lVert E_i - \hat{E}_i \rVert_1$, where $E_i$ and $\hat{E}_i$ denote the RMS energy features of the $i$-th frame from the reference and generated audio ~\cite{choi2023foley}. For subjective evaluation, we conduct a 20-minute listening test with nine participants using a 5-point Likert scale. Two key dimensions are evaluated: Audio Quality, which measures perceptual naturalness, and Time Alignment, which measures synchronization with the conditioning signal. All test samples are provided on the demo page. For each sample, we present the reference audio together with the waveform and RMS energy curves, allowing participants to assess temporal synchronization through both listening and visual inspection.

\section{Results}
\subsection{Objective and Subjective Evaluation}
The left panel of Table~\ref {tab:1} presents the objective results for DCASE and FSD-Kaggle2018. Overall, the proposed DP-DiT and DP-U-Net variants achieve strong performance across all metrics with substantially fewer parameters than baselines. Specifically, DP-U-Net obtains the lowest E-L1 and FAD scores, indicating superior temporal alignment and perceptual quality, while DP-U-Net Small (3.26M) remains highly competitive. The performance gap between DP-U-Net and DP-DiT further underscores the architectural advantage of U-Net over DiT for waveform generation. Among the baselines, Mamba-Foley outperforms T-Foley, suggesting the importance of latent space design. Diffwave achieves relatively low E-L1 but suffers from high FAD scores, indicating limited perceptual fidelity. Conversely, TF-DiT performs poorly, confirming that direct spectrogram modeling along the temporal axis is inadequate. In summary, our models benefit from more effective inductive biases, enabling substantial parameter reduction while maintaining strong temporal accuracy and perceptual quality. 

Subjective evaluation results are summarized in the right panel of Table~\ref{tab:1}. Our proposed DP-U-Net consistently achieves the highest scores in both Audio Quality and Temporal Alignment, significantly outperforming all baselines. Notably, even the DP-U-Net Small variant outperforms much larger models such as T-Foley and Mamba-Foley, demonstrating that our DP architecture can deliver high-fidelity, well-synchronized Foley audio with minimal parameters.

\subsection{Model Complexity and Efficiency Evaluation}
\begin{table}[t]
\centering
\caption{Inference efficiency and model complexity.}
\label{tab:inference_efficiency}
\begin{tabular}{lccc}
\toprule
\textbf{Method}
& \textbf{RTF} $\downarrow$
& \textbf{GFLOPs} $\downarrow$
& \textbf{Memory (GB)}
\\
\midrule
T-Foley \cite{chung2024t}                 & 2.61 & 54.15 & 2.11 \\
Mamba-Foley \cite{colombo2025mambafoley}  & 4.05 & 64.67 &  0.44 \\
DiffWave \cite{kongdiffwave}              & 6.24 & 1302.84 & 9.74 \\
TF-DiT                                   & 6.85 & 9.58 & 0.12 \\
\midrule
\textbf{DP-DiT}                                & 31.06 & 349.53 & 1.12 \\
\textbf{DP-U-Net}                                 & 4.37 & 196.78 & 1.22 \\
\textbf{DP-U-Net Small}                          & 3.20 & 78.38 & 1.15 \\
\bottomrule
\end{tabular}
\vspace{-1em} 
\end{table}

\begin{figure}[!tb]
  \centering
  \centerline{\includegraphics[width=9cm]{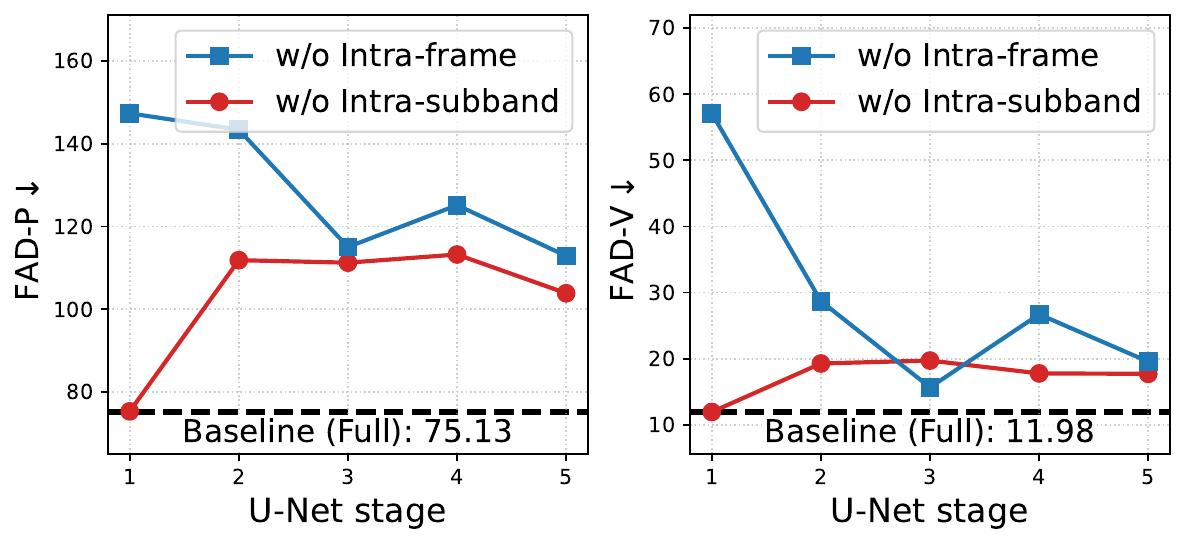}}
 % \vspace{2.0cm}
  % \centerline{(a) Result 1}\medskip
\caption{Layer-wise ablation on the 5-stage DP-U-Net Small by independently omitting a single attention block at each stage.}
\label{fig:4}
\vspace{-1em} 
\end{figure}

% \subsection{Case Study on Modeling Frequency Domain Features}
\begin{figure*}[!tb]
  \centering
  \centerline{\includegraphics[width=15.5cm]{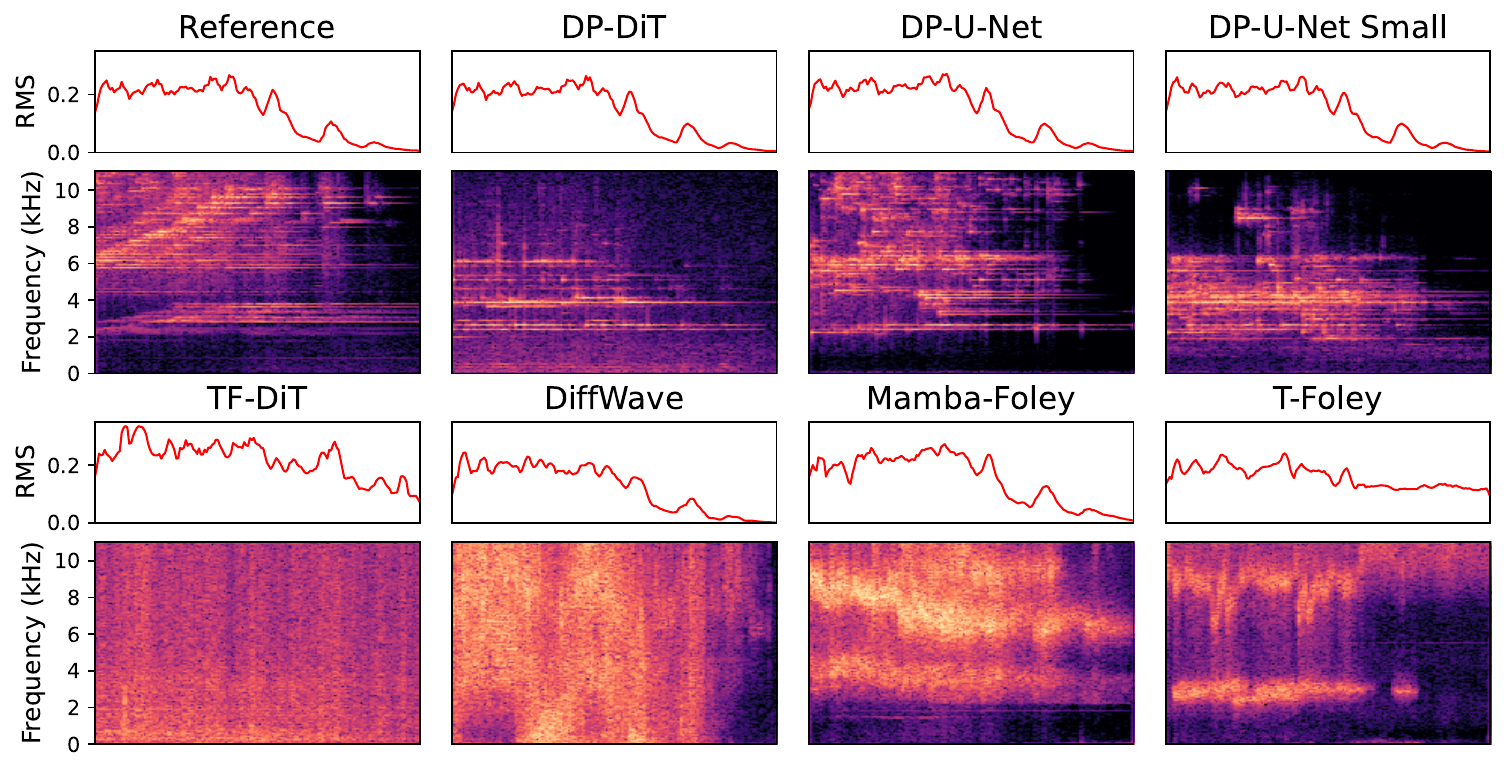}}
 % \vspace{2.0cm}
  % \centerline{(a) Result 1}\medskip
\caption{Reference and generated RMS energy curves and audio examples for the Chime class from the FSD-Kaggle2018 dataset.}
\label{fig:5}
\vspace{-2em} 
\end{figure*}

We assess efficiency using Real-Time Factor (RTF) under 200 diffusion steps with CFG on 4-second clips, peak GPU memory, and GFLOPs per forward pass. All of the tests are conducted on a single NVIDIA V100. As shown in Table~\ref{tab:inference_efficiency}, DP-U-Net maintains high generation quality while keeping inference cost moderate, and DP-U-Net Small further reduces computational complexity. In contrast, DP-DiT exhibits substantially higher RTF and GFLOPs due to its depth. Among baselines, T-Foley achieves the fastest inference speed, Mamba-Foley is the most memory-efficient, and DiffWave incurs the highest computational cost. Although TF-DiT requires relatively low computation, its limited generation quality restricts its applicability.

\subsection{Ablation Study and Case Study}
To evaluate the contribution of each component, we conduct a layer-wise ablation study on the trained 5-stage DP-U-Net Small model. Specifically, we independently remove either the intra-frame (spectral) attention or the intra-subband (temporal) attention at each stage, and measure the resulting performance changes to quantify their individual impact. As shown in Figure~\ref{fig:4}, both modules contribute to the generation quality, with the Intra-frame attention providing a larger contribution, underscoring the importance of direct frequency modeling.

Figure~\ref{fig:5} presents a qualitative case study on the Chime sound from the FSD-Kaggle2018 dataset. All samples are generated using the same temporal conditioning derived from the reference audio and the same random seed. While all methods can track the provided RMS conditioning signal, significant differences emerge in the generated spectrograms. Specifically, only the proposed TF-domain DP models: DP-DiT, DP-U-Net, and DP-U-Net Small successfully generate meaningful and complex frequency-domain features that align with the reference, highlighting the importance of architectural design for modeling complex audio spectral patterns.

\section{Conclusion}
We propose an efficient waveform diffusion model for Foley sound generation based on a fully attention-based U-Net with dimension-wise self-attention. By introducing more effective inductive biases for temporal–spectral modeling, the proposed approach achieves superior generation quality on the DCASE and FSD-Kaggle2018 datasets using only a few million parameters, demonstrating that high-quality Foley sound generation can be achieved with significantly improved efficiency through well-designed architectures.

\printbibliography

@article{ghose2022foleygan,
  title={Foleygan: Visually guided generative adversarial network-based synchronous sound generation in silent videos},
  author={Ghose, Sanchita and Prevost, John J},
  journal={IEEE Transactions on Multimedia},
  volume={25},
  pages={4508--4519},
  year={2022},
  publisher={IEEE}
}

@techreport{xie2023x,
  title={The X-LANCE system for DCASE2023 challenge task 7: Foley sound synthesis track b},
  author={Xie, Zeyu and Xu, Xuenan and Li, Baihan and Wu, Mengyue and Yu, Kai},
  year={2023},
  institution={Tech. Rep., June}
}

@inproceedings{qi2024mtdiffusion,
  title={Mtdiffusion: multi-task diffusion model with dual-unet for foley sound generation},
  author={Qi, Anbin and Xie, Xiang and Wang, Jing},
  booktitle={ICASSP 2024-2024 IEEE International Conference on Acoustics, Speech and Signal Processing (ICASSP)},
  pages={461--465},
  year={2024},
  organization={IEEE}
}

@inproceedings{huang2025rhythmic,
  title={Rhythmic foley: A framework for seamless audio-visual alignment in video-to-audio synthesis},
  author={Huang, Zhiqi and Luo, Dan and Wang, Jun and Liao, Huan and Li, Zhiheng and Wu, Zhiyong},
  booktitle={ICASSP 2025-2025 IEEE International Conference on Acoustics, Speech and Signal Processing (ICASSP)},
  pages={1--5},
  year={2025},
  organization={IEEE}
}

@article{yuan2023text,
  title={Text-driven foley sound generation with latent diffusion model},
  author={Yuan, Yi and Liu, Haohe and Liu, Xubo and Kang, Xiyuan and Wu, Peipei and Plumbley, Mark D and Wang, Wenwu},
  journal={arXiv preprint arXiv:2306.10359},
  year={2023}
}

@inproceedings{chung2024t,
  title={T-foley: A controllable waveform-domain diffusion model for temporal-event-guided foley sound synthesis},
  author={Chung, Yoonjin and Lee, Junwon and Nam, Juhan},
  booktitle={ICASSP 2024-2024 IEEE International Conference on Acoustics, Speech and Signal Processing (ICASSP)},
  pages={6820--6824},
  year={2024},
  organization={IEEE}
}

@article{liu2023audioldm,
  title={Audioldm: Text-to-audio generation with latent diffusion models},
  author={Liu, Haohe and Chen, Zehua and Yuan, Yi and Mei, Xinhao and Liu, Xubo and Mandic, Danilo and Wang, Wenwu and Plumbley, Mark D},
  journal={arXiv preprint arXiv:2301.12503},
  year={2023}
}

@inproceedings{colombo2025mambafoley,
  title={Mambafoley: Foley sound generation using selective state-space models},
  author={Colombo, Marco Furio and Ronchini, Francesca and Comanducci, Luca and Antonacci, Fabio},
  booktitle={ICASSP 2025-2025 IEEE International Conference on Acoustics, Speech and Signal Processing (ICASSP)},
  pages={1--5},
  year={2025},
  organization={IEEE}
}

@inproceedings{pascual2023full,
  title={Full-band general audio synthesis with score-based diffusion},
  author={Pascual, Santiago and Bhattacharya, Gautam and Yeh, Chunghsin and Pons, Jordi and Serr{\`a}, Joan},
  booktitle={ICASSP 2023-2023 IEEE International Conference on Acoustics, Speech and Signal Processing (ICASSP)},
  pages={1--5},
  year={2023},
  organization={IEEE}
}

@inproceedings{kongdiffwave,
  title={DiffWave: A Versatile Diffusion Model for Audio Synthesis},
  author={Kong, Zhifeng and Ping, Wei and Huang, Jiaji and Zhao, Kexin and Catanzaro, Bryan},
  booktitle={International Conference on Learning Representations}
}

@article{van2016wavenet,
  title={Wavenet: A generative model for raw audio},
  author={Van Den Oord, Aaron and Dieleman, Sander and Zen, Heiga and Simonyan, Karen and Vinyals, Oriol and Graves, Alex and Kalchbrenner, Nal and Senior, Andrew and Kavukcuoglu, Koray and others},
  journal={arXiv preprint arXiv:1609.03499},
  volume={12},
  pages={1},
  year={2016}
}

@inproceedings{ku2025generative,
  title={Generative speech foundation model pretraining for high-quality speech extraction and restoration},
  author={Ku, Pin-Jui and Liu, Alexander H and Korostik, Roman and Huang, Sung-Feng and Fu, Szu-Wei and Juki{\'c}, Ante},
  booktitle={ICASSP 2025-2025 IEEE International Conference on Acoustics, Speech and Signal Processing (ICASSP)},
  pages={1--5},
  year={2025},
  organization={IEEE}
}

@article{iashchenko2023undiff,
  title={UnDiff: Unsupervised voice restoration with unconditional diffusion model},
  author={Iashchenko, Anastasiia and Andreev, Pavel and Shchekotov, Ivan and Babaev, Nicholas and Vetrov, Dmitry},
  journal={arXiv preprint arXiv:2306.00721},
  year={2023}
}

@inproceedings{moliner2023solving,
  title={Solving audio inverse problems with a diffusion model},
  author={Moliner, Eloi and Lehtinen, Jaakko and V{\"a}lim{\"a}ki, Vesa},
  booktitle={ICASSP 2023-2023 IEEE International Conference on Acoustics, Speech and Signal Processing (ICASSP)},
  pages={1--5},
  year={2023},
  organization={IEEE}
}

@inproceedings{peebles2023scalable,
  title={Scalable diffusion models with transformers},
  author={Peebles, William and Xie, Saining},
  booktitle={Proceedings of the IEEE/CVF international conference on computer vision},
  pages={4195--4205},
  year={2023}
}

@inproceedings{luo2020dual,
  title={Dual-path rnn: efficient long sequence modeling for time-domain single-channel speech separation},
  author={Luo, Yi and Chen, Zhuo and Yoshioka, Takuya},
  booktitle={ICASSP 2020-2020 IEEE International Conference on Acoustics, Speech and Signal Processing (ICASSP)},
  pages={46--50},
  year={2020},
  organization={IEEE}
}

@article{choi2023foley,
  title={Foley sound synthesis at the dcase 2023 challenge},
  author={Choi, Keunwoo and Im, Jaekwon and Heller, Laurie and McFee, Brian and Imoto, Keisuke and Okamoto, Yuki and Lagrange, Mathieu and Takamichi, Shinosuke},
  journal={arXiv preprint arXiv:2304.12521},
  year={2023}
}

@article{loshchilov2017decoupled,
  title={Decoupled weight decay regularization},
  author={Loshchilov, Ilya and Hutter, Frank},
  journal={arXiv preprint arXiv:1711.05101},
  year={2017}
}

@article{kilgour2018fr,
  title={Fr$\backslash$'echet audio distance: A metric for evaluating music enhancement algorithms},
  author={Kilgour, Kevin and Zuluaga, Mauricio and Roblek, Dominik and Sharifi, Matthew},
  journal={arXiv preprint arXiv:1812.08466},
  year={2018}
}

@article{fonseca2018general,
  title={General-purpose tagging of freesound audio with audioset labels: Task description, dataset, and baseline},
  author={Fonseca, Eduardo and Plakal, Manoj and Font, Frederic and Ellis, Daniel PW and Favory, Xavier and Pons, Jordi and Serra, Xavier},
  journal={arXiv preprint arXiv:1807.09902},
  year={2018}
}

@inproceedings{nguyen2024fregrad,
  title={Fregrad: Lightweight and fast frequency-aware diffusion vocoder},
  author={Nguyen, Tan Dat and Kim, Ji-Hoon and Jang, Youngjoon and Kim, Jaehun and Chung, Joon Son},
  booktitle={ICASSP 2024-2024 IEEE International Conference on Acoustics, Speech and Signal Processing (ICASSP)},
  pages={10736--10740},
  year={2024},
  organization={IEEE}
}

@article{liu2024audioldm,
  title={Audioldm 2: Learning holistic audio generation with self-supervised pretraining},
  author={Liu, Haohe and Yuan, Yi and Liu, Xubo and Mei, Xinhao and Kong, Qiuqiang and Tian, Qiao and Wang, Yuping and Wang, Wenwu and Wang, Yuxuan and Plumbley, Mark D},
  journal={IEEE/ACM Transactions on Audio, Speech, and Language Processing},
  volume={32},
  pages={2871--2883},
  year={2024},
  publisher={IEEE}
}

@inproceedings{lu2021study,
  title={A study on speech enhancement based on diffusion probabilistic model},
  author={Lu, Yen-Ju and Tsao, Yu and Watanabe, Shinji},
  booktitle={2021 Asia-Pacific Signal and Information Processing Association Annual Summit and Conference (APSIPA ASC)},
  pages={659--666},
  year={2021},
  organization={IEEE}
}

@inproceedings{shi2026unsupervised,
  title={Unsupervised Single-Channel Audio Separation with Diffusion Source Priors},
  author={Shi, Runwu and Li, Chang and Wang, Jiang and Zhang, Rui and Khan, Nabeela and Yen, Benjamin and Ashizawa, Takeshi and Nakadai, Kazuhiro},
  booktitle={Proceedings of the AAAI Conference on Artificial Intelligence},
  volume={40},
  number={30},
  pages={25348--25356},
  year={2026}
}

\end{document}